\documentclass[twocolumn,aps,superscriptaddress,showpacs,floatfix,prc]{revtex4}
\UseRawInputEncoding
\usepackage{mathrsfs}
\usepackage{amssymb}
\usepackage{amsmath}
\usepackage{graphicx}
\usepackage[normalem]{ulem}
\usepackage[dvips]{color}
\usepackage{bm}
\usepackage{longtable}\usepackage{slashed}

\usepackage{newtxtext}
\usepackage{enumitem}

\usepackage[varg]{txfonts}

\usepackage{epstopdf}

\usepackage{times}

\renewcommand\sout{\bgroup \color{red} \ULdepth=-.5ex \ULset}

\renewcommand{\rm}[1]{\textrm{#1}}
\renewcommand{\d}{\mathrm{d}}

\begin{document}

\title{Intrinsic Correlations Among Characteristics of
Neutron-rich Matter\\
Imposed by the Unbound Nature of Pure Neutron Matter}

\author{Bao-Jun Cai\footnote{bjcai87@gmail.com}}
\affiliation{Quantum Machine Learning Laboratory, Shadow Creator Inc., Shanghai 201208, China}
\author{Bao-An Li\footnote{Bao-An.Li$@$tamuc.edu}}
\affiliation{Department of Physics and Astronomy, Texas A$\&$M
University-Commerce, Commerce, TX 75429-3011, USA}
\date{\today}

\begin{abstract}
\noindent
{\bfseries Background:} 
The equation of state (EOS) $E(\rho,\delta)$ of neutron-rich nucleonic matter at density $\rho$ and isospin asymmetry $\delta$ can be approximated as the sum of the symmetric nuclear matter (SNM) EOS $E_0(\rho)$ and a power series in $\delta^2$ with the coefficient $E_{\rm{sym}}(\rho)$ of its first term the so-called nuclear symmetry energy.  Great efforts have been devoted to studying the characteristics of 
both the SNM EOS $E_0(\rho)$ and the symmetry energy $E_{\rm{sym}}(\rho)$ using various experiments and theories over the last few decades. While much progresses have been made in constraining parameters characterizing 
the $E_0(\rho)$ and $E_{\rm{sym}}(\rho)$ around the saturation density $\rho_0$ of SNM, such as the incompressibility $K_0$ of SNM EOS as well as the magnitude $S$ and slope $L$ of nuclear symmetry, the parameters characterizing
the high-density behaviors of both $E_0(\rho)$ and $E_{\rm{sym}}(\rho)$, such as the  skewness $J_0$ and kurtosis $I_0$ of SNM EOS as well as the curvature $K_{\rm{sym}}$ and skewness $J_{\rm{sym}}$ of the symmetry energy are still poorly known. Moreover, most attention has been put on constraining the characteristics of the SNM EOS and the symmetry energy separately as if the $E_0(\rho)$ and $E_{\rm{sym}}(\rho)$  are completely independent. 
\\\\
{\bfseries Purpose:} Since the EOS of PNM is the sum of the SNM EOS and all symmetry energy coefficients in expanding the $E(\rho,\delta)$ as a power series in $\delta^2$,
the unbound nature of PNM requires intrinsic correlations between the SNM EOS parameters and those characterizing the symmetry energy independent of any nuclear many-body theory.  
We investigate these intrinsic correlations and their applications in better constraining the poorly known high-density behavior of nuclear symmetry energy. 
 \\\\
{\bfseries Method:} We first derive an expression for the saturation density ${\rho_{\rm{sat}}(\delta)}$ of neutron-rich matter up to order $\delta^6$ in terms of the EOS parameters. Setting ${\rho_{\rm{sat}}(\delta)}$ =0 for PNM as required by its unbound nature leads to a sum rule for the EOS parameters. We then analyze this sum rule at different orders in $\delta^2$ to find approximate expressions of the high-density symmetry energy parameters in terms of 
the relatively better determined SNM EOS parameters and the symmetry energy around $\rho_0$.
\\\\
{\bfseries Results:} Several novel correlations connecting the characteristics of SNM EOS with those of nuclear symmetry energy are found. In particular, at the lowest-order of approximations, 
the bulk parts of the slope $L$, curvature $K_{\rm{sym}}$ and skewness $J_{\rm{sym}}$ of the symmetry energy are found to be $L\approx K_0/3, K_{\rm{sym}}\approx LJ_0/2K_0$ and $J_{\rm{sym}}\approx I_0L/3K_0$, respectively. 
High-order corrections to these simple relations can be written in terms of the small ratios of high-order EOS parameters. The resulting intrinsic correlations among some of the EOS parameters reproduce very nicely their relations predicted by various microscopic nuclear many-body theories and phenomenological models constrained by available data of terrestrial experiments and astrophysical observations in the literature. 
\\\\
{\bfseries Conclusion:} The unbound nature of PNM is fundamental and the required intrinsic correlations among the EOS parameters characterizing both the SNM EOS and symmetry energy are universal. 
These intrinsic correlations provide a novel and model-independent tool not only for consistency checks but also for investigating the poorly known high-density properties of neutron-rich matter by using those with smaller uncertainties.

\end{abstract}

\maketitle

%\tableofcontents

\section{Introduction}\label{S1}
The equation of state (EOS) of cold asymmetric nucleonic matter (ANM) given in terms of the energy per nucleon $E(\rho,\delta)$ at density $\rho$ and isospin asymmetry $\delta\equiv (\rho_{\rm{n}}-\rho_{\rm{p}})/\rho$ between neutron and proton densities $\rho_{\rm{n}}$ and $\rho_{\rm{p}}$ is a fundamental quantity in nuclear physics and a basic input for various applications especially in astrophysics\,\cite{Rin80}.
The $E(\rho,\delta)$ is often expanded around $\delta=0$ to further define the EOS $E_0(\rho)$ of symmetric nucleonic matter (SNM) and various orders of the so-called nuclear symmetry energy according to 
\begin{equation}\label{eos1}
E(\rho,\delta)=E_0(\rho)+\sum_{i=1}^{\infty} E_{\rm{sym},2i}(\rho )\delta ^{2i}.
\end{equation}
Setting $\delta=1$, the above equation reduces to an approximate relation among the EOS $E_{\rm{PNM}}(\rho)\equiv E(\rho,1)$ of pure neutron matter (PNM), the SNM EOS $E_0(\rho)$ and all of the symmetry energy terms, i.e., 
$
E_{\rm{PNM}}(\rho)=E_0(\rho)+\sum_{i=1}^{\infty} E_{\rm{sym},2i}(\rho ).
$
Normally the expansion in $\delta$ ends at the quadratic term with $i=1$ in the so-called parabolic approximation, and one generally refers the $E_{\rm{sym},2}(\rho)\equiv E_{\rm{sym}}(\rho)$ as the symmetry energy by neglecting all higher order terms. One then expands the SNM EOS $E_0(\rho)$ and the symmetry energy $E_{\rm{sym}}(\rho)$  around the saturation density $\rho_0$ of SNM with their characteristic coefficients, e.g., the incompressibility $K_0$, skewness $J_0$ and kurtosis $I_0$ of SNM as well as the magnitude $S$, slope $L$, curvature $K_{\rm{sym}}$ and skewness $J_{\rm{sym}}$ of the symmetry energy $E_{\rm{sym}}(\rho)$.  

Much efforts in both experiments and theories have been devoted to investigating the characteristics of SNM during the last four decades and those of the symmetry energy over the last two decades, respectively. Indeed, much progresses have been made in constraining both the $E_0(\rho)$ and nuclear symmetry energy especially around and below $\rho_0$, see, e.g, Refs.\,\cite{Dan02,ditoro,LCK08,Tesym,Col14,Bal16,Oer17,Garg18} for reviews. 
On the other hand, much progresses have also been made in recent years in understanding properties of low-density PNM using the state-of-the-art microscopic nuclear many-body theories and advanced computational techniques, e.g., chiral effective field theories\,\cite{Heb15} and quantum Monte Carlo techniques\,\cite{Car15}. All calculations indicate that both the energy and pressure in PNM approach zero smoothly and monotonically as the density vanishes, reflecting the unbound nature of PNM\,\cite{FP,MS,Sch05,Gez10,Hut20}. Moreover, it has been shown that the unitary Fermi gas EOS $E_{\rm{UG}}(\rho)=\xi E_{\rm{FG}}(\rho)$ in terms of the EOS of a free Fermi gas for neutrons $E_{\rm{FG}}(\rho)$
and a Bertsch parameter $\xi\approx0.376$\,\cite{UG} provides a lower bound to the $E_{\rm{PNM}}(\rho)$ at low densities\,\cite{Tew17}, demonstrating vividly the deep quantum nature of the system in the so-called contact region\,\cite{Gio08}. 

Imposing constraints on theories or fitting model predictions with experimental observables will naturally introduce correlations among the EOS parameters. Indeed, some interesting correlations have been found especially among the characteristics of either the SNM EOS $E_0(\rho)$ or the symmetry energy $E_{\rm{sym}}(\rho)$.  It is also interesting to note that the low-density PNM EOS from the microscopic theories has been used as a boundary condition to calibrate some phenomenological models\,\cite{Fa1,Fa2,PPNP} and to explore some correlations among the EOS parameters\,\cite{Newton12,Newton20}. In particular, the condition that the energy per neutron in PNM vanishes at zero density naturally leads to a linear correlation between $K_{\rm{sym}}$  and $3S$--$L$\,\cite{Mar19}. However, most of the correlations among the EOS parameters found so far are very model dependent especially those involving the high-order coefficients, see, e.g., Refs.\,\cite{Tew17,Mar19,Maz13,Lida14,Pro14,Col14,Mon17,Hol18,LiBA2020,Han20}. Since these correlations are known to have significant effects on properties of both nuclei and neutron stars, see, e.g., Refs.\,\cite{Maz13,Newton12,Li19}, a better understanding of the correlations among the EOS parameters have significant ramifications in both nuclear physics and astrophysics. 

In this work, we show that the unbound nature of PNM alone, especially its vanishing pressure $P$ at zero density (i.e., the saturation density of ANM approaches zero as $\delta\rightarrow 1$), naturally leads to a sum rule linking intrinsically the EOS parameters independent of any theory. Analyses of this sum rule at different orders of the expansions in $\delta$ and $\rho$ lead to novel correlations relating the characteristics of SNM EOS with those of nuclear symmetry energy. In particular, at the lowest-order of approximations, we found that $L\approx K_0/3, K_{\rm{sym}}\approx LJ_0/2K_0$ and $J_{\rm{sym}}\approx I_0L/3K_0$, respectively. Corrections to these simple relations by considering high-order terms reproduce nicely the empirical correlations among some of the EOS parameters reported previously in the literature. 

The rest of the paper is organized as follows. In section \ref{S2}, we recall the basic definitions of ANM EOS parameters. 
Intrinsic correlations among the EOS parameters and their general implications are given in section \ref{S3}. 
We then discuss potential applications of the intrinsic correlations.  
As the first example, we give in section \ref{S4} a scheme for estimating the slope $L$ of nuclear symmetry energy. Section \ref{S5} is devoted to investigating the correlation between the curvature $K_{\rm{sym}}$ of the symmetry energy and the $L,K_0$ and $J_0$. Relevant comparisons with the empirical $K_{\rm{sym}}$--$L$ relations in the literature will also be given in this section.
In section \ref{S6b} and section \ref{S6a}, two direct implications of the $K_{\rm{sym}}$--$L$ relation on the correlation between $L$ and the symmetry energy magnitude $S$ at $\rho_0$ as well as the isospin-dependent part of the incompressibility coefficient $K_{\rm{sat,2}}$ will be studied. Section \ref{S7} gives a possible constraint on the skewness $J_{\rm{sym}}$ of the symmetry energy.
A short summary is finally given in section \ref{S8}.

\section{Universal Constraints on the EOS parameters of Neutron-rich Matter by the Unbound  Nature of Pure Neutron Matter}

\subsection{Characteristics of Isospin Asymmetric Nuclear Matter}\label{S2}

For easy of our discussions in the following, we summarize here the necessary notations and recall some terminologies we adopted. 
Using the $\delta$ and $\chi=(\rho-\rho_0)/3\rho_0$ as two perturbative variables, the EOS of ANM can be expanded around SNM at $\rho_0$ generally as $E(\rho,\delta)=w_{ij}\delta^{2i}\chi^j$ where the repeated indices are summed over, with each term characterized by $\delta^{2i}\chi^j$. In the following, we call ``$2i+j$'' the order of the quantity considered\,\cite{Cai17x}.
In this sense, $E_0(\rho_0)=w_{00}$ is the only zeroth-order term from $i=j=0$, and the first-order term is absent due to the vanishing pressure at $\rho_0$ in SNM by definition of its saturation point.
At second order, we have $K_0=9\rho_0^2E''_0(\rho_0)=2w_{02}$ as well as $S\equiv E_{\rm{sym}}(\rho_0)=w_{10}$. Very similarly, we have at third order $L=3\rho_0E'_{\rm{sym}}(\rho_0)=w_{11}$ and $J_0=27\rho_0^3E'''_0(\rho_0)=6w_{03}$. While at fourth order, we have $I_0=81\rho_0^4E''''_0(\rho_0)=24w_{04},K_{\rm{sym}}=9\rho_0^2E''_{\rm{sym}}(\rho_0)=2w_{12}$ and $S_4\equiv E_{\rm{sym,4}}(\rho_0)=w_{20}$. Finally, the skewness $J_{\rm{sym}}=27\rho_0^3E'''_{\rm{sym}}(\rho_0)=6w_{13}$ of the symmetry energy $E_{\rm{sym}}(\rho)$ and the slope coefficient $L_{\rm{sym,4}}=3\rho_0E'_{\rm{sym,4}}(\rho_0)=w_{21}$ of the fourth-order symmetry energy $E_{\rm{sym,4}}(\rho)$ are both at order five.

\subsection{Intrinsic Correlations of EOS parameters and Their General Implications}\label{S3}
The saturation density $\rho_{\rm{sat}}(\delta)$ for ANM with isospin asymmetry $\delta$ is defined as the point where the pressure vanishes, namely $P(\rho_{\rm{sat}}(\delta))=0$, or equivalently $
{\partial E(\rho,\delta)}/{\partial\rho}|_{\rho=\rho_{\rm{sat}}(\delta)}=0$. After a long but straightforward calculation by expanding all terms in Eq.\,(\ref{eos1}) as functions of $\chi$ and the saturation density $\rho_{\rm{sat}}$ as a function of $\delta$ (see the Appendix in Ref.\,\cite{Che09}), one can obtain
\begin{equation}\label{PNM-p-order}
{\rho_{\rm{sat}}(\delta)}/{\rho_0}\approx1+\Psi_2\delta^2+\Psi_4\delta^4+\Psi_6\delta^6 +\mathcal{O}(\delta^8),
\end{equation}
with
\begin{align}
\Psi_2=&-\frac{3L}{K_0},\\
\Psi_4=&\frac{3K_{\rm{sym}}L}{K_0^2}-\frac{3L_{\rm{sym,4}}}{K_0}
-\frac{3J_0L^2}{2K_0^3},\\
\Psi_6=&\left(\frac{3K_{\rm{sym}}L}{K_0^2}-\frac{3J_0L^2}{2K_0^3}
-\frac{3L_{\rm{sym,4}}}{K_0}\right)\left(\frac{J_0L}{K_0^2}-\frac{K_{\rm{sym}}}{K_0}\right)\notag\\
&+\frac{I_0L^3}{2K_0^4}
-\frac{3J_{\rm{sym}}L^2}{2K_0^3}+\frac{3K_{\rm{sym,4}}L}{K_0^2}
-\frac{3L_{\rm{sym,6}}}{K_0},
\end{align}
here $K_{\rm{sym,4}}$ is the curvature of the fourth-order symmetry energy and $L_{\rm{sym,6}}$ is the slope of the sixth-order symmetry energy.
The expressions for $\Psi_2$ and $\Psi_4$ were first given in Ref.\,\cite{Che09}.

Required by the unbound nature of PNM as shown by all existing nuclear many-body theories, the saturation density of ANM eventually decreases from $\rho_0$ to zero as $\delta$ goes from zero (SNM) to 1 (PNM).
Quantitatively, to the order $\delta^6$ this boundary condition requires $1+\Psi_2+\Psi_4+\Psi_6=0$.  It can be rewritten using the characteristic EOS coefficients as an approximate sum rule
\begin{align}\label{def-L-order}
1=&\frac{3L}{K_0}-\left(\frac{3K_{\rm{sym}}L}{K_0^2}-\frac{3J_0L^2}{2K_0^3}
+\frac{3L_{\rm{sym,4}}}{K_0}\right)\left(1+\frac{J_0L}{K_0^2}+\frac{K_{\rm{sym}}}{K_0}\right)\notag\\
&-\frac{I_0L^3}{2K_0^4}
+\frac{3J_{\rm{sym}}L^2}{2K_0^3}-\frac{3K_{\rm{sym,4}}L}{K_0^2}
+\frac{3L_{\rm{sym,6}}}{K_0}.
\end{align}
Clearly, it establishes an intrinsic correlation among the EOS parameters. Compared to the equation one would obtain from setting the energy per neutron $E_{\rm{PNM}}(0)=0$ at zero density, through couplings between isospin-independent and isospin-dependent coefficiens the above relation provides a more stringent constraint for the high-order EOS parameters without involving the binding energy $E_0(\rho_0)$ of SNM and the magnitude $S$ of the symmetry energy at $\rho_0$. In fact, the latter has been well determined to be around $S=E_{\rm{sym}}(\rho_0)=31.7 \pm 3.2$~MeV by extensive analyses of terrestrial experiments and astrophysical observations \cite{LiBA13,Oer17} as well as {\it ab initio} nuclear-many theory predictions \cite{Ohio20}. 

The analytic expressions for the saturation density at different orders of $\delta^2$ characterized by $\Psi_2,\Psi_4$ and $\Psi_6$, etc., are very useful since they fundamentally encapsulate the intrinsic correlations among the characteristic EOS parameters. In this sense, intrinsic equations at different orders of $\delta^2$ could be obtained.
Specifically, if one truncates the saturation density $\rho_{\rm{sat}}^{\rm{PNM}}/\rho_0$ in PNM to order $\delta^2$, an intrinsic equation $1+\Psi_2=0$ is obtained.
More accurately this should be an intrinsic inequality $0\lesssim1+\Psi_2\lesssim 1$. However, it is known that at the lowest orders of $\delta^2$ in the so-called parabolic approximation, 
the $\rho_{\rm{sat}}^{\rm{PNM}}$ is not necessarily zero in some of the many-body calculations, see examples given in Ref.\,\cite{LiBA98}. Nevertheless, the Eq.\,(\ref{PNM-p-order}) provides a scheme to improve gradually the 
accuracy of calculating the saturation density of ANM. For example, an intrinsic equation $1+\Psi_2+\Psi_4=0$ or $1+\Psi_2+\Psi_4+\Psi_6=0$ could be obtained if the $\rho_{\rm{sat}}^{\rm{PNM}}/\rho_0$ is truncated at order $\delta^4$ or $\delta^6$, respectively. As the order of truncation increases, the quantity $1+\sum_{j=1}\Psi_{2j}=1+\Psi_2+\Psi_4+\cdots$ becomes more and more close to zero, and the intrinsic correlations or the sum rule $1+\sum_{j=1}\Psi_{2j}=0$ obtained becomes more and more accurate. Moreover, as the accuracy in estimating the ${\rho_{\rm{sat}}(\delta)}$ increase, more high-order EOS parameters get involved as one expects. 
 
We point out the following two basic usages of the intrinsic equations (at different orders of $\delta^2$):
\begin{enumerate}[leftmargin=*]
\item The intrinsic equations can be effectively used to establish useful connections between parameters characterizing the SNM EOS and those describing the symmetry energy. For example, one can immediately obtain from the intrinsic equation $1+\Psi_2=0$ at order $\delta^2$ the relation $L\approx K_0/3$, and if the coefficient $K_0$ is better constrained then the $L$ coefficient could by subsequently inferred. Moreover, if both of them are independently determined, this simple relation allows for a consistency check. Naturally, if higher order contributions are included, this simple relation is expected to be modified. In section \ref{S4}, we investigate this issue in more details. The most important physics outcome is that the approximate relations, e.g., $L\approx K_0/3$, provide a useful guide for developing phenomenological models and microscopic theories.
\item As mentioned above,  as the truncation order of the expansion (\ref{PNM-p-order}) increases, more and more characteristic coefficients would be included, i.e., they will emerge in the intrinsic equations as the order increases. Then, the correlations can potentially allow one to extract high-order (poorly known) coefficients from the lower-order (better known) ones. An example given in section \ref{S5} on extracting the $K_{\rm{sym}}$ from its correlation with $L$ will demonstrate this point in detail.
\end{enumerate}

Moreover, a few interesting points related to Eq.\,(\ref{PNM-p-order}) and Eq.\,(\ref{def-L-order}) are worth emphrasing:
\begin{enumerate}[leftmargin=*]
\item  Since the saturation density $\rho_{\rm{sat}}(\delta)/\rho_0$ is obtained order by order, one naturally expects the higher order term contributes less. If the $\rho_{\rm{sat}}(\delta)/\rho_0$ is truncated at order $\delta^2$, then one has $\Psi_2=-1$. Actually what we have obtained is $0\gtrsim\Psi_2\gtrsim-1$. Next if the $\Psi_4$ contribution is taken into consideration, we have $\rho_{\rm{sat}}^{\rm{PNM}}/\rho_0=1+\Psi_2+\Psi_4=0$, i.e., $\Psi_2=-1-\Psi_4$. Simultaneously setting $|\Psi_2|\lesssim1$ and $|\Psi_4|\lesssim|\Psi_2|$ gives the constraint on $\Psi_4$ as $-1/2\lesssim\Psi_4\lesssim0$.
When the $\Psi_6$ term is included, a similar analysis on the value of $\Psi_6$ could be made.

\item  Although equation (\ref{def-L-order}) implies a complicated intrinsic correlation among all the parameters involved, certain terms could directly be found to be less important. For example, the characteristic coefficients related to the fourth-order symmetry energy $E_{\rm{sym,4}}(\rho)$, namely $L_{\rm{sym,4}}$ and $K_{\rm{sym,4}}$ are expected to be small, since the value of $S_4\equiv E_{\rm{sym,4}}(\rho_0)$ is known to be smaller than about 2\,MeV\,\cite{Cai12,Gon17,PuJ17,Lee98}.
Particularly, by adopting a density dependence similar to the symmetry energy $E_{\rm{sym}}(\rho)$\,\cite{Rus16} as $E_{\rm{sym,4}}(\rho)=S_4(\rho/\rho_0)^{\gamma}$ with $\gamma$ an effective density-dependence parameter, one obtains $L_{\rm{sym,4}}=3\gamma S_4$ and $K_{\rm{sym,4}}\approx 9\gamma(\gamma-1)S_4$. A conservative estimate with $0\lesssim\gamma\lesssim2$ indicates that $0\,\rm{MeV}\lesssim L_{\rm{sym,4}}\lesssim12\,\rm{MeV}$ and $-4.5\,\rm{MeV}\lesssim K_{\rm{sym,4}}\lesssim36\,\rm{MeV}$. Consequently, one can safely neglect the term ``$-3K_{\rm{sym,4}}L/2K_0^3$'' in Eq.\,(\ref{def-L-order}) since its magnitude is smaller than 1.5\% with the currently known most probable values of $L$\,\cite{LiBA13,Oer17} and $K_0$ \cite{Garg18}. For similar reasons, the last term in Eq.\,(\ref{def-L-order}) involving the slope $L_{\rm{sym,6}}$ of the six-order symmetry energy $E_{\rm{sym,6}}(\rho)$ can also be neglected. However, the term involving the $L_{\rm{sym,4}}$ might be important and it is kept in the following discussions.

\item By truncating Eq.\,(\ref{PNM-p-order}) at different orders of $\delta^2$, different intrinsic correlation equations are obtained. But they are not all independent. 
As the truncation order increases, the intrinsic correlation becomes more general since more and more characteristic coefficients are taken into consideration. Of course, the resulting dependences/correlations among the EOS parameters look gradually more complicated.
\end{enumerate}

\section{Applications of the Intrinsic Correlations among EOS parameters}\label{S0}
In this section, we present a few applications of the sum rule in Eq.\,(\ref{def-L-order}) at different orders of $\delta^2$. To justify some of our numerical approximations and ease of our discussions, it is useful to note here the currently known ranges of the lower-order parameters and the rough magnitudes of the high-order parameters. In particular, we shall use the empirical values of $K_0\approx240\pm 20\,\rm{MeV}$\,\cite{Garg18,You99,Shl06,Che12,Col14}, $J_0\approx-300\,\rm{MeV}$\,\cite{Cai17x} and $I_0\approx -146\pm 1728$\,MeV\,\cite{XieLi-JPG} for the SNM EOS. For the symmetry energy, it is known that  $L\approx 60\pm30\,\rm{MeV}$\,\cite{LiBA13}, $K_{\rm{sym}}\approx-80\,\rm{MeV}$\,\cite{Xie20} and $J_{\rm{sym}}\approx300\,\rm{MeV}$\,\cite{Mon17}. While the community has not reached a consensus on the exact ranges of some of the high-order parameters, the reference values enable us to make rough estimates and judge if some of the ratios/terms in our analyses can be neglected. 

\subsection{Estimating the Slope $L$ of Nuclear Symmetry Energy}\label{S4}
As the first application of the intrinsic equations, we derive an expression for the slope $L$ of the symmetry energy in terms of better known quantities corrected by some ratios of other parameters that are small.
The main purpose of this analysis is to show how the unbound nature of PNM naturally gives a good estimate for $L$. 

As discussed in section \ref{S3}, the intrinsic equation at order $\delta^2$ gives the lowest-order approximation $L\approx K_0/3$. Thus, by introducing the dimensionless quantity $
x={3L}/{K_0}$ (consequently $x^{(0)}\approx 1$ corresponding to $L\approx K_0/3$ could be found) and the following combinations
\begin{align}
\psi_0=&1-\frac{3L_{\rm{sym,4}}}{K_0}-\frac{3L_{\rm{sym,6}}}{K_0},\\
\psi_1=&1-\frac{K_{\rm{sym}}}{K_0}+\frac{L_{\rm{sym,4}}J_0}{K_0^2}-\frac{K_{\rm{sym,4}}}{K_0}+\frac{K_{\rm{sym}}^2}{K_0^2},\\
\psi_2=&1-\frac{3K_{\rm{sym}}}{K_0}+\frac{J_{\rm{sym}}}{J_0},\\\psi_3=&1-\frac{3J_0^2}{K_0I_0},
\end{align}
we obtain the equation for $x$ by rewriting Eq.\,(\ref{def-L-order}) as
\begin{equation}\label{def-cubic-x}
\psi_0-\psi_1x-\frac{1}{6}\frac{J_0}{K_0}\psi_2x^2+\frac{1}{54}\frac{I_0}{K_0}\psi_3x^3=0.
\end{equation}
Rewriting the Eq.\,(\ref{def-L-order}) in the form (\ref{def-cubic-x}) is mainly for the convenience of discussing its relevance for estimating $L$.
If on the other hand the interested quantity is $K_{\rm{sym}}$, then Eq.\,(\ref{def-L-order}) could be cast into the form $A K_{\rm{sym}}^2+B K_{\rm{sym}}+C=0$ with $A,B$ and $C$ being some relevant coefficients independent of $K_{\rm{sym}}$. We note that the $J_0$ appearing in $\psi_1$ comes from the $\Psi_4$, while the $J_{\rm{sym}}$ and $I_0$ appearing in $\psi_2$ and $\psi_3$ come from the $\Psi_6$.

Different approximations of Eq.\,(\ref{def-cubic-x}) and analysis could be developed. In the following, we discuss approximate solutions of Eq.\,(\ref{def-cubic-x}) at $\delta^2, \delta^4$ and $\delta^6$ order, separately.
At the $\delta^2$ order, besides the simplest estimation $L\approx K_0/3$ by taking $\psi_0\approx\psi_1\approx1$ in Eq.\,(\ref{def-cubic-x}) and simultaneously neglecting the $x^2$ and $x^3$ terms, one can also find that ${L}/{K_0}>0$, i.e., the sign of the incompressibility $K_0$ is the same as that of the slope $L$. The latter can be seen from the relation 
${\rho_{\rm{sat}}(\delta)}/{\rho_0}\approx1+\Psi_2\delta^2+\mathcal{O}(\delta^4)$ and $\Psi_2=-{3L}/{K_0}$. As the isospin asymmetry $\delta$ increases from zero to a small finite value, the saturation density has to be reduced. Thus, $\Psi_2$ has to be negative, requiring ${L}/{K_0}>0$. 

Systematically, we can generalize the relation $L\approx K_0/3$ as
\begin{equation}\label{pq-1}
L\approx \frac{K_0}{3}\times\Bigg(1+\mbox{``higher order corrections''}\Bigg),
\end{equation}
when the higher order terms $\Psi_4\delta^4$, $\Psi_6\delta^6$, etc., are taken into account in the saturation density $\rho_{\rm{sat}}/\rho_0$.
For example, at $\delta^4$ order, by considering the $\Psi_4$ term, but simultaneously neglecting the $\Psi_6$ term, the coefficients related to the fourth-order symmetry energy as well as the small terms $(K_{\rm{sym}}/K_0)^2$ and 
$J_0K_{\rm{sym}}/2K_0^2$,  one can obtain 
\begin{equation}\label{eeq-1}
\frac{1}{6}\frac{J_0}{K_0}x^2+\left(1-\frac{K_{\rm{sym}}}{K_0}\right)x-1=0.
\end{equation}
Its solution leads to 
\begin{align}
L=&\left(1-\frac{K_{\rm{sym}}}{K_0}\right)\frac{K_0^2}{J_0}
\left[\sqrt{1+\frac{2}{3}\frac{J_0}{K_0}\left(1-\frac{K_{\rm{sym}}}{K_0}\right)^{-2}}-1\right]\label{kc-0},\\
\approx&\frac{K_0}{3}\times\left(1-\frac{K_{\rm{sym}}}{K_0}\right)^{-1}\left[1-\frac{1}{6}\frac{J_0}{K_0}\left(1-\frac{K_{\rm{sym}}}{K_0}\right)^{-2}\right]\label{kc-1},
\end{align}
where the second approximation is obtained by noticing that $|1+(2J_0/3K_0)(1-K_{\rm{sym}}/K_0)^{-2}|\ll1$ using the empirical values of the involved parameters given earlier.
In addition, the positiveness of the discriminant of Eq.\,(\ref{eeq-1}) limits the skewness $J_0$ to $J_0\gtrsim -3K_0(1-K_{\rm{sym}}/K_0)^2/2$.
Since $K_{\rm{sym}}/K_0\approx -1/3$ and $J_0/6K_0\approx -5/24 $ are both small and at the same level, the above expression for $L$ can be further approximated as
\begin{equation}
L\approx \frac{K_0}{3}\times\left(1+\frac{K_{\rm{sym}}}{K_0}-\frac{1}{6}\frac{J_0}{K_0}\right)\label{LO4}.
\end{equation}

Similarly, at $\delta^6$ order, again by neglecting the coefficients related to the fourth-order and the sixth-order symmetry energies as well as the other small terms, the solution of Eq.\,(\ref{def-cubic-x}) leads to 
\begin{align}\label{L-app}
L\approx \frac{K_0}{3}\times &\left[1+\frac{K_{\rm{sym}}}{K_0}-\frac{J_0+J_{\rm{sym}}}{6K_0}\right.\notag\\
&\left.+\frac{I_0}{54K_0}\left(1+\frac{4K_{\rm{sym}}}{K_0}-\frac{5J_0}{6K_0}+\frac{I_0}{18K_0}\right)\right].
\end{align}
While the kurtosis $I_0$ and the incompressibility $K_0$ have similar orders of magnitude, the second line of (\ref{L-app}) contributes only about 2\% compared to the leading contribution ``1''.
Using the empirical values of the EOS parameters given earlier, the ``high order contributions'' in (\ref{L-app}) is estimated to be about $-33$\%. Consequently, the $L$ is about $53.3\,\rm{MeV}$, compared to the simplest estimation of about 80\,MeV from $L\approx K_0/3$. 
Moreover, the coefficient $1/54$ together with the small ratios enable us to further approximate the expression (\ref{L-app}) to 
\begin{equation}\label{LO6}
L\approx \frac{K_0}{3}\times\left(1+\frac{K_{\rm{sym}}}{K_0}-\frac{1}{6}\frac{J_0}{K_0}-\frac{1}{6}\frac{J_{\rm{{sym}}}}{K_0}+\frac{1}{54}\frac{I_0}{K_0}\right).
\end{equation}
As noticed before, the $J_{\rm{sym}}$ and $I_0$ appear only in the $\Psi_6$. Neglecting them, the above expression naturally reduces to Eq.\,(\ref{LO4}) valid at the $\delta^4$ order as one expects. 

Our analyses above indicate clearly that as the truncation order of $\rho_{\rm{sat}}^{\rm{PNM}}$ increases, the higher order terms become eventually irrelevant although they appear in a complicated manner.
To understand the results intuitively, it is useful to look at the analogy with the period $T$ of small angle oscillations of a simple pendulum 
\begin{equation}
T(\theta)\approx2\pi\sqrt{\frac{l}{g}}\times\left(1+\frac{1}{16}\theta^2+\frac{11}{3072}\theta^4\right),
\end{equation}
where $\theta$ is the maximum angle and $l$ is the length of the pendulum. For $\theta=1$, we obtain from the above expression the period $T(1)\approx 1.066\times T(0)$. While the exact result is $T(1)\approx 1.085\times T(0)$, indicating that although the apparent perturbation element $\theta$ is not much smaller than 1, the perturbative expansion is still very effective (useful) near $\theta\approx1$ due to the small in-front coefficients $1/16$ and $11/3072$ in the expansion of $T(\theta)$.

\subsection{Intrinsic Correlation between $K_{\rm{sym}}$ and $L$}\label{S5}

 \renewcommand*\figurename{\small Fig.}
\begin{figure*}
\centering
\includegraphics[height=4.1cm]{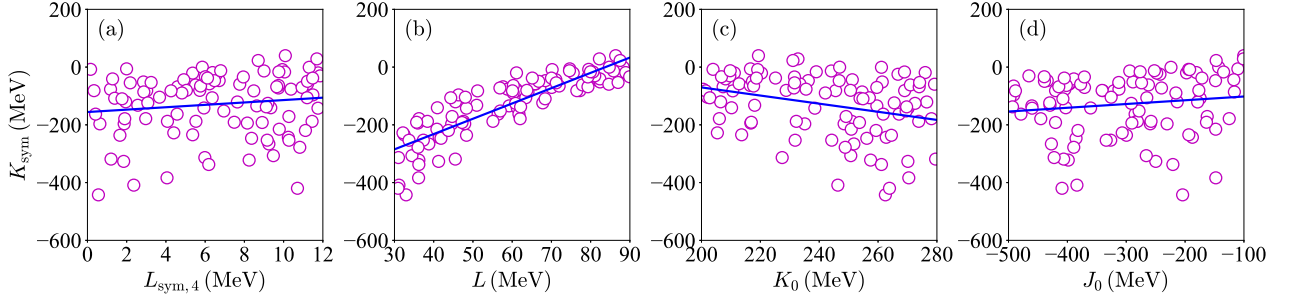}
 \caption{(Color Online). Correlations between $K_{\rm{sym}}$ and $L_{\rm{sym,4}},L,K_0$ and $J_0$ according to Eq.\,(\ref{def_estKsym}).}\label{fig_simulateKsymL-o1}
\end{figure*}

Besides estimating the slope $L$, the intrinsic equations could also be used to investigate the correlations among certain EOS parameters. As $K_{\rm{sym}}$ first appears at $\delta^4$ order in the $\Psi_4$ term, 
through the intrinsic equation $1+\Psi_2+\Psi_4=0$, we obtain the following relation for $K_{\rm{sym}}$,
\begin{equation}\label{def_estKsym}
K_{\rm{sym}}\approx K_0\left(1-\frac{1}{3}\frac{K_0}{L}+\frac{1}{2}\frac{J_0}{K_0}\frac{L}{K_0}+\frac{L_{\rm{sym,4}}}{L}\right).
\end{equation}
One can first check the magnitude of $K_{\rm{sym}}$ using this equation. By taking $K_0\approx 240\,\rm{MeV},J_0\approx-300\,\rm{MeV},L\approx60\,\rm{MeV}$ and $L_{\rm{sym,4}}\approx6\,\rm{MeV}$, one obtains $K_{\rm{sym}}\approx-93\,\rm{MeV}$. In addition, if one assumes $|\Psi_4/\Psi_2|\lesssim1$, then one finds $-253\,\rm{MeV}\lesssim K_{\rm{sym}}\lesssim227\,\rm{MeV}$. These results are consistent with the constraint on $K_{\rm{sym}}$ obtained recently from Bayesian analyses of neutron star observations\,\cite{Xie20,Xie19}.

It is necessary to point out that the relations (\ref{kc-0}) or (\ref{kc-1}) and  (\ref{def_estKsym}) are not independent, and in fact they are effectively two different presentations of the same intrinsic equation.
However, they are usefully different in the sense that they give separately the expressions for $L$ and  $K_{\rm{sym}}$ in terms of their main parts plus small corrections determined by the ratios of other EOS parameters. 
The analysis above indicates that the $K_{\rm{sym}}$ is closely correlated with $K_0,L$ and $J_0$ while the $L_{\rm{sym,4}}/L$ has negligible effects as we discussed earlier. 
Moreover, taking the lowest-order approximation for $L$, i.e.,  $L\approx K_0/3$, the expression (\ref{def_estKsym}) can be further reduced to 
\begin{equation}
 K_{\rm{sym}}\approx K_0\left(\frac{1}{2}\frac{J_0}{K_0}\frac{L}{K_0}+\frac{L_{\rm{sym,4}}}{L}\right)\approx LJ_0/2K_0.
 \end{equation}
 One then immediately finds qualitatively that $K_{\rm{sym}}$ correlates positively with $L$ and $J_0$ but negatively with $K_0$. 

In the following, we investigate more quantitatively the correlations of $K_{\rm{sym}}$ with other EOS parameters using the more accurate relation (\ref{def_estKsym}). For this purpose, we perform a Monte Carlo sampling of the EOS parameters in their currently known ranges. Shown in Fig.\,\ref{fig_simulateKsymL-o1} are the correlations between $K_{\rm{sym}}$--$L_{\rm{sym,4}}$, $K_{\rm{sym}}$--$L$, $K_{\rm{sym}}$--$K_0$ and $K_{\rm{sym}}$--$J_0$, respectively. In this study, the parameters are randomly sampled simultaneously and uniformly in the range of $0\,\rm{MeV}\lesssim L_{\rm{sym,4}}\lesssim  12\,\rm{MeV},30\,\rm{MeV}\lesssim  L\lesssim 90\,\rm{MeV},200\,\rm{MeV}\lesssim K_0\lesssim 280\,\rm{MeV}$ and $-500\,\rm{MeV}\lesssim  J_0\lesssim -100\,\rm{MeV}$.
From the results shown we clearly observe that the $K_{\rm{sym}}$ has weak positive correlations with both $L_{\rm{sym,4}}$ and $J_0$ but a strong positive (negative) correlation with $L$ ($K_0$).
When more experimental constraints on the EOS parameters become available, the accuracy and robustness of the correlations are expected to be improved. However, the correlation patterns revealed here should stay the same.

The strengths of correlations between the $K_{\rm{sym}}$ and the other EOS parameters can be quantified using the quantity $\Theta(\phi_i)=
\delta\phi_i{\partial K_{\rm{sym}}}/{\partial\phi_i}$ where $\phi_i=L_{\rm{sym,4}},L,K_0,J_0$. More specifically, we find that $\Theta(L)\approx258\,\rm{MeV}$, $\Theta(L_{\rm{sym,4}})\approx32\,\rm{MeV},\Theta(K_0)\approx-120\,\rm{MeV}$, and $\Theta(J_0)\approx50\,\rm{MeV}$, respectively. These numbers clearly quantify the strengths of the correlations shown in Fig.\,\ref{fig_simulateKsymL-o1}.  It is interesting to note here that the results of our model-independent analyses are very consistent with the findings of Ref.\,\cite{Mar19} (see Fig.\,7 there). In the latter, the correlations were obtained from some basic physical constraints imposed on the Taylor's expansion of
the binding energy in ANM around $\rho_0$\,\cite{Mar19}. One of the constraints Ref.\,\cite{Mar19} used is that the EOS of PNM at zero density is zero. The present analysis thus shares with Ref.\,\cite{Mar19} the requirement that the PNM is unbound. However, as we pointed out earlier, the vanishing pressure of PNM puts a more stringent constraints on the high-order EOS parameters than its vanishing binding energy at zero density. 

The near-linear correlations between $K_{\rm{sym}}$ and the other EOS parameters can also be described more quantitatively. For example, the correlation between $K_{\rm{sym}}$
and $L$ can be written as $K_{\rm{sym}}\approx a_1L+b_1$.  By minimizing the algebraic error, the coefficients $a_1$ and $b_1$ can be obtained  as
\begin{align}
a_1\approx\frac{\langle K_{\rm{sym}}L\rangle-\langle K_{\rm{sym}}\rangle\langle L\rangle}{\langle L^2\rangle-\langle L\rangle^2},
\end{align}
and $b_1=\langle K_{\rm{sym}}\rangle-a_1\langle L\rangle$, here  the average $\langle\cdots\rangle$ for one independent simulation with total $m$ points is defined simply as $\langle k\rangle=m^{-1}\sum_{i=1}^mk^{(i)}$.
By independently sampling $n$ times one obtains the standard uncertainty of $a_1$ or $b_1$ as
\begin{equation}
\sigma_f=
\sqrt{
\frac{1}{n}\sum_{i=1}^nf^{(i),2}-\left(\frac{1}{n}\sum_{i=1}^nf^{(i)}\right)^2
},
\end{equation}
where $f^{(i)}\leftrightarrow a_1^{(i)},b^{(i)}_1$. Using $m=10^2,n=10^4$ in our simulations, we find $\overline{a}_1=n^{-1}\sum_{i=1}^na^{(i)}_1\approx5.25\pm0.40$ and similarly $\overline{b}_1=n^{-1}\sum_{i=1}^nb^{(i)}_1\approx-441\pm28\,\rm{MeV}$. Taking $L\approx60\,\rm{MeV}$ in $K_{\rm{sym}}\approx\overline{a}_1L+\overline{b}_1$ gives $K_{\rm{sym}}\approx-126\,\rm{MeV}$.
For the correlation between $K_{\rm{sym}}$ and $L_{\rm{sym,4}}$, namely $K_{\rm{sym}}\approx a_2L_{\rm{sym,4}}+b_2$, one has $\overline{a}_2\approx4.42\pm3.11$ and $\overline{b}_2\approx-152\pm22\,\rm{MeV}$, similarly for  $K_{\rm{sym}}$ and $K_0$, namely $K_{\rm{sym}}\approx a_3K_0+b_3$, one has $\overline{a}_3\approx-1.66\pm0.44$ and $\overline{b}_3\approx272\pm103\,\rm{MeV}$. Finally for the correlation between $K_{\rm{sym}}$ and $J_0$, i.e., $K_{\rm{sym}}\approx a_4J_0+b_4$, the results are found to be $\overline{a}_4\approx0.13\pm0.09$ and $\overline{b}_4\approx-88\pm30\,\rm{MeV}$.
It is necessary to point out that  while the central values of $\overline{a}_i$ and $\overline{b}_i$ (with $i=1\sim4$) will approach those determined by the central values of the characteristic coefficients, e.g., $L$ and $K_0$, for the simulation as $n\to\infty$ according to the law of large numbers\,\cite{Blu20}, the magnitude of the uncertainty is affected by the choice of $m$.
A smaller $m$ leads to a larger uncertainty as one expects.

\renewcommand*\figurename{\small Fig.}
\begin{figure}[h!]
\centering
\includegraphics[height=4.1cm]{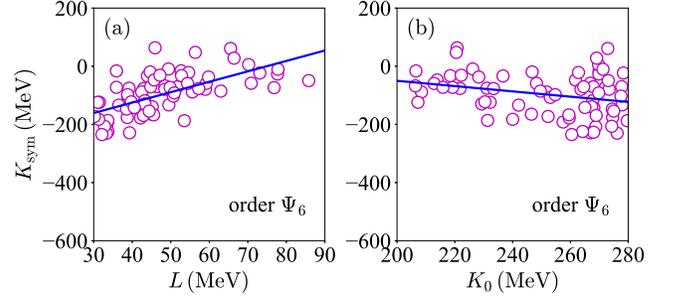} \caption{(Color Online). Correlations between $K_{\rm{sym}}$ and $L$ (left) and that between $K_{\rm{sym}}$ and $K_0$ (right) to order $\Psi_6$.}\label{fig_simulateKsymL-o2}
\end{figure} 

When high order contributions are considered, we can similarly study the correlation between the $K_{\rm{sym}}$ and the other EOS coefficients. For instance, at the $\delta^6$ order, the relevant equation for calculations is obtained by neglecting the coefficients $L_{\rm{sym,6}}$ and $K_{\rm{sym,4}}$ in Eq.\,(\ref{def-L-order}).
Fig.\,\ref{fig_simulateKsymL-o2} shows results of a Monte Carlos sampling of the correlations at $\delta^6$ order using $0\,\rm{MeV}\lesssim  J_{\rm{sym}}\lesssim 2000\,\rm{MeV}$ and $-2000\,\rm{MeV}\lesssim  I_0\lesssim 2000\,\rm{MeV}$ as well as those ranges given earlier for the low-order EOS parameters.
It is seen that the positive (negative) correlation between $K_{\rm{sym}}$ and $L$ ($K_0$) is unchanged when the relevant higher order contributions are included.
This means that the qualitative features of the intrinsic correlations obtained earlier from the Eq.\,(\ref{def-L-order}) at the $\delta^4$ order are kept and stable, while the fitting coefficients are affected quantitatively. More specifically, now for the correlation $K_{\rm{sym}}\approx a_1'L+b_1'$ the $n$-average of $a_1'$ and $b_1'$ are found to be $\overline{a}_1'\approx3.61\pm0.43$ and $\overline{b}_1'\approx-268\pm21\,\rm{MeV}$, respectively.  Similarly for the correlation $K_{\rm{sym}}\approx a_3'K_0+b_3'$, we have $\overline{a}_3'\approx-0.75\pm0.30$ and $\overline{b}_3'\approx90\pm71\,\rm{MeV}$, respectively.

\renewcommand*\figurename{\small Fig.}
\begin{figure}[h!]
\centering
\includegraphics[height=6.2cm]{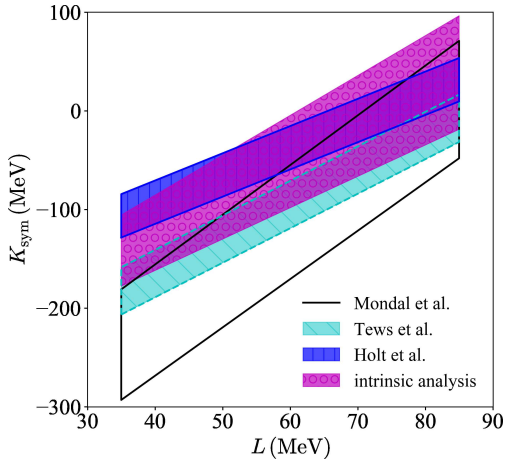}
 \caption{(Color Online). Correlation between $K_{\rm{sym}}$ and $L$ from intrinsic analysis as well as the prediction on it from Refs.\,\cite{Tew17,Mon17,Hol18}.}\label{fig_Ksym-L-corr}
\end{figure}

It is also interesting to compare our $K_{\rm{sym}}$--$L$ correlation with typical results obtained from other approaches in the literature\,\cite{Tew17,Mon17,Hol18}. More specifically, Ref.\,\cite{Mon17} gave a correlation $K_{\rm{sym}}\approx (-4.97\pm0.07)(3S-L)+67\pm2\,\rm{MeV}$ at 68\% confidence level from analyzing over 520 predictions of the Skyrme--Hartree--Fock energy density functionals and the relativistic mean field theories. Their result using $S\approx 32\pm3\,\rm{MeV}$ is shown with the black lines in Fig.\,\ref{fig_Ksym-L-corr}. The authors of Ref.\,\cite{Tew17} did a similar analysis as Ref.\,\cite{Mon17} but applied additional constraints. Their result $K_{\rm{sym}}\approx 3.50L-306\pm24\,\rm{MeV}$ is shown as the sky blue band in Fig.\,\ref{fig_Ksym-L-corr}. In Ref.\,\cite{Hol18},  a $K_{\rm{sym}}$--$L$ correlation was derived by using the Fermi liquid theory with its parameters calibrated by the chiral effective field theory at sub-saturation densities. Their correlation $K_{\rm{sym}}\approx 2.76L-203\pm22\,\rm{MeV}$ is shown as the cyan band.  While our result is shown with the purple band. 
It is seen that our correlation is highly consistent with the results from the three other studies. They all overlap largely around the upper boundary from the analysis of Ref.\,\cite{Mon17}. More quantitatively, taking $L\approx60\,\rm{MeV}$ in $K_{\rm{sym}}\approx\overline{a}_1'L+\overline{b}_1'$ gives $K_{\rm{sym}}\approx-52\,\rm{MeV}$, while the same value of $L$ leads to a mean value of $K_{\rm{sym}}\approx-112\,\rm{MeV},-96\,\rm{MeV}$ and $-37\,\rm{MeV}$ using the constraints from Refs.\,\cite{Mon17,Tew17,Hol18}, respectively.
We also notice that a very recent study based on the so-called KIDS energy functional\,\cite{Han20} gave a $K_{\rm{sym}}$--$L$ correlation consistent with the results discussed above. More quantitatively, they concluded that the curvature $K_{\rm{sym}}$ is probably negative and not lower than $-200$\,MeV\,\cite{Han20}.
Finally, it is also interesting to note that the analysis combining the tidal deformability of neutron stars via a $\chi^2$-based covariance approach also showed that the coefficients $K_{\rm{sym}}$ and $L$ are strongly correlated\,\cite{Mal20,Mal18}.

\subsection{Implications of the Intrinsic $K_{\rm{sym}}$--$L$ Correlation on the Incompressibility of Neutron-rich Matter along its Saturation Line}
\label{S6a}
The incompressibility of neutron-rich matter 
\begin{equation}
K_{\rm{sat}}(\delta)\equiv\left. 9\rho^2_{\rm{sat}}\frac{\partial^2E(\rho,\delta)}{\partial\rho^2}\right|_{\rho=\rho_{\rm{sat}}}\approx K_0+K_{\rm{sat,2}}\delta^2+\mathcal{O}(\delta^4)
\end{equation}
is an important quantity directly related to the ongoing studies of various collective modes and stability of neutron-rich nuclei\,\cite{Garg18}. 
The strength of its isospin-dependent part can be written as\,\cite{Che09}
\begin{equation}
K_{\rm{sat,2}}=K_{\rm{sym}}-6L-J_0L/K_0.
\end{equation}
By using Eq.\,(\ref{def_estKsym}) for $K_{\rm{sym}}$, the latter can be rewritten as
\begin{equation}
K_{\rm{sat,2}}\approx \left(1+\frac{L_{\rm{sym,4}}}{L}
-\frac{K_0}{3L}\right)K_0-\left(6+\frac{J_0}{2K_0}\right)L.
\end{equation}
Numerically, we find $K_{\rm{sat,2}}\approx-378\,\rm{MeV}$ using the empirical values of EOS parameters given earlier.
This value is consistent with the latest study on $K_{\rm{sat,2}}$ within the KIDS framework where it was found that the $K_{\rm{sat,2}}$ should roughly lie between $-400$\,MeV and $-300$\,MeV\,\cite{Han20}.
Similar to what are shown in Fig.\,\ref{fig_simulateKsymL-o1}, using the above expression for $K_{\rm{sat,2}}$ one can also analyze its correlations with $K_0$, $L,L_{\rm{sym,4}}$ and $J_0$, separately.
Moreover, noticing that $|L_{\rm{sym,4}}/L|\ll1$ and $L\approx K_0/3$, the $K_{\rm{sat,2}}$ can be further approximated as 
\begin{equation}
K_{\rm{sat,2}}\approx -2K_0-J_0/6.
\end{equation}
This relation clearly demonstrates that the uncertainty of $K_{\rm{sat,2}}$ mainly comes from the poorly known $J_0$ although its contribution has a shrinking factor of $-1/6$.

\subsection{Implications of the Intrinsic $K_{\rm{sym}}$--$L$ Correlation on the $L$--$S$ Correlation of the Symmetry Energy}
\label{S6b}
The near-linear correlation between $K_{\rm{sym}}$ and $L$ also has an important implication on the correlation between $L$ and $S$ of the symmetry energy at $\rho_0$. Noticing that at an arbitrary density $\rho$
\begin{equation}\label{S-L-a}
\frac{\d L(\rho)}{\d E_{\rm{sym}}(\rho)}
=\left.\frac{\d L(\rho)}{\d\rho}\right/\frac{\d E_{\rm{sym}}(\rho)}{\d\rho}=3+\frac{K_{\rm{sym}}(\rho)}{L(\rho)}
\end{equation}
according to the basic definitions of the characteristic coefficients of symmetry energy $E_{\rm{sym}}(\rho)$. 
By taking $\rho=\rho_0$, the relation (\ref{S-L-a}) gives $\d L/\d S=3+K_{\rm{sym}}/L$. For instance, the $L$ and $K_{\rm{sym}}$ in the free Fermi gas model are $L=2S$ and $K_{\rm{sym}}=-2S=-L$, respectively, automatically fulfilling the relation (\ref{S-L-a}).
Using the intrinsic correlation $K_{\rm{sym}}\approx aL+b$ found earlier and by integrating $\int\d S=\int\d L(3+K_{\rm{sym}}/L)^{-1}=\int\d L(a+3+b/L)^{-1}$, we obtain the following relation between $S$ and $L$
\begin{equation}\label{S-L-for}
S(L)=\frac{1}{a+3}\left(L-\frac{b}{a+3}\ln\left[(a+3)L+b\right]\right)
+\rm{const.},
\end{equation}
where the constant is determined via some reference point $(S_{\rm{H}},L_{\rm{H}})$, e.g., $S_{\rm{H}}\approx32\,\rm{MeV}$ and $L_{\rm{H}}\approx 60\,\rm{MeV}$ at $\rho_0$ according to the surveys of 
available data\,\cite{LiBA13,Oer17}.

\renewcommand*\figurename{\small Fig.}
\begin{figure}[h!]
\centering
\includegraphics[height=6.2cm]{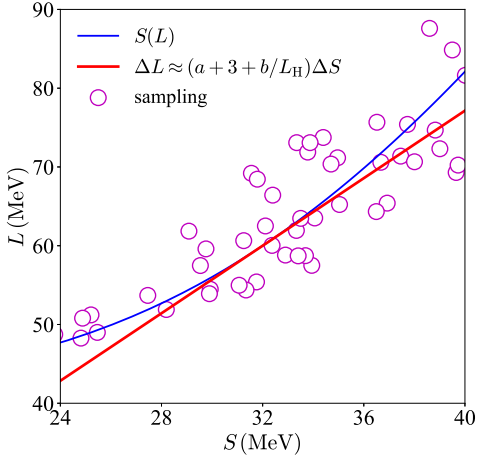}
 \caption{(Color Online). Correlation between $L$ and $S$.}\label{fig_S-Lcorr}
\end{figure} 

Shown in Fig.\,\ref{fig_S-Lcorr} with the pink circles are the Monte Carlo samplings of the $L$--$S$ correlation within the empirical EOS parameter ranges and adopting $\overline{a}_1'\approx3.61$ and $\overline{b}_1'\approx-268\,\rm{MeV}$ from the $\delta^6$ order analysis discussed earlier. Due to the near-linear correlation between $K_{\rm{sym}}$ and $L$, the correlation between $L$ and $S$ is also found to be near-linear, although it is not so obvious that the relation (\ref{S-L-for}) is linear. Since the near-linearity between $K_{\rm{sym}}$ and $L$ is intrinsic (only the fitting coefficients vary if the intrinsic equation is truncated at different orders), the near-linearity between $L$ and $S$ is also expected to be intrinsic. Specifically, one can easily show that 
\begin{equation}\label{S-L-for1}
\Delta L\approx\Phi\Delta S\left[1-\left({b}/{2L_{\rm{H}}^2}\right)\Delta S\right],~~\Phi=a+3+{b}/{L_{\rm{H}}},
\end{equation}
where $\Delta L=L-L_{\rm{H}},\Delta S=S-S_{\rm{H}}$, and $\Phi$ is simply the value of $3+K_{\rm{sym}}/L$ taken at $L_{\rm{H}}$.
Numerically, we obtain the relation $L\approx 2.14\Delta S+0.08\Delta S^2+L_{\rm{H}}$ by using $a=\overline{a}_1'\approx3.61$ and $b=\overline{b}_1'\approx-268\,\rm{MeV}$. 
The quadratic correction $0.08\Delta S^2$ is small, thus $\Delta L\approx 2.14\Delta S$ and consequently $L\approx2.14S-8.6\,\rm{MeV}$,  which is very close to the free Fermi gas model prediction on the $L$--$S$ relation $L=2S$.
The resulting linearized $\Delta L$--$\Delta S$ correlation (\ref{S-L-for1}) (by neglecting the $\Delta S^2$ correction) and the full $L$--$S$ correlation  (\ref{S-L-for}) are shown with the red and blue lines, respectively, in Fig.\,\ref{fig_S-Lcorr}.  Around the currently known most probable value of $S=E_{\rm{sym}}(\rho_0)=31.7 \pm 3.2$\,MeV\,\cite{LiBA13,Oer17}, both the $L$--$S$ and the Monte Carlo samplings are approximately linear consistently. 
Moreover, by putting the $L$ in terms of $S$ back into the relation $K_{\rm{sym}}\approx aL+b$, we obtain $
K_{\rm{sym}}\approx 7.74S-299\,\rm{MeV}$, which has a large deviation from its free Fermi gas model counterpart, see also Ref.\,\cite{Hol18}.

\subsection{Estimating the $K_{\rm{sym}}$--$J_{\rm{sym}}$ Correlation}\label{S7}
The $J_{\rm{sym}}$ first emerges in $\Psi_6$, one can thus investigate its intrinsic correlations with the other EOS parameters starting from the order $\delta^6$. Since it is a high-order parameter, its value is poorly known and some of 
its correlations especially with the low-order parameters are expected to be weak. Solving the intrinsic equation (\ref{def-L-order}) by neglecting $L_{\rm{sym,4}},K_{\rm{sym,4}}$ and $L_{\rm{sym,6}}$ leads to the following estimation for $J_{\rm{sym}}$
\begin{align}\label{def_estJsym}
J_{\rm{sym}}\approx&\frac{2K_0^3}{3L^2}\left(1-\frac{3L}{K_0}\right)\notag\\
&+\left(\frac{2K_0K_{\rm{sym}}}{L}-J_0\right)
\left(1+\frac{J_0L}{K_0^2}-\frac{K_{\rm{sym}}}{K_0}\right)
+\frac{I_0L}{3K_0}.
\end{align}
Taking the lowest-order approximations for $L\approx K_0/3$ and $K_{\rm{sym}}\approx LJ_0/2K_0$ obtained earlier, the above equation can be further reduced to $J_{\rm{sym}}\approx I_0L/3K_0$. 
Numerically, we have $J_{\rm{sym}}\approx281\,\rm{MeV}$ by putting the empirical values of $K_0, J_0, I_0, L$ and $K_{\rm{sym}}$ given earlier into Eq.\,(\ref{def_estJsym}). This value is consistent with the constrain 
$J_{\rm{sym}}\approx 296.8\pm73.6\,\rm{MeV}$ from analyzing the systematics of over 520 energy density functionals in Ref.\,\cite{Mon17}. It is also consistent with the $J_{\rm{sym}}=90\pm 334$ MeV at 68\% confidence level from a recent Bayesian analysis in Ref.\,\cite{Som20} (see Tab.\,4 there).

\begin{figure}[h!]
\centering
\includegraphics[height=6.2cm]{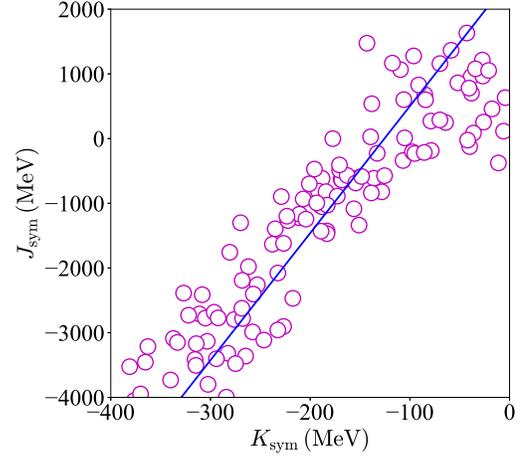}
 \caption{(Color Online). Correlation between $J_{\rm{sym}}$ and $K_{\rm{sym}}$ according to the relation (\ref{def_estJsym}).}\label{fig_simulate-Jsym-Ksym}
\end{figure}

Fig.\,\ref{fig_simulate-Jsym-Ksym} shows the $J_{\rm{sym}}$--$K_{\rm{sym}}$ correlation from our Monte Carlo samplings in the range of $-400\,\rm{MeV}\lesssim K_{\rm{sym}}\lesssim 0\,\rm{MeV}$ while the other EOS 
parameters are taken randomly in their empirical ranges given earlier. It is seen that there is a strong correlation between $J_{\rm{sym}}$ and $K_{\rm{sym}}$.
In fact, the $J_{\rm{sym}}$ depends on the $K_{\rm{sym}}$ quadratically in Eq.\,(\ref{def_estJsym}). Obviously, the uncertainty of $J_{\rm{sym}}$ is very large and its strong dependence on the still poorly constrained $K_{\rm{sym}}$ partially explains why constraining the  $J_{\rm{sym}}$ is difficult. Moreover, the last term in (\ref{def_estJsym}), namely $I_0L/3K_0\approx I_0/12$ also contributes significantly to the uncertainty of $J_{\rm{sym}}$ since we have little knowledge on $I_0$.

\renewcommand*\figurename{\small Fig.}

One can obtain the fitting parameters $p$ and $q$ appearing in the linear fit $J_{\rm{sym}}\approx pK_{\rm{sym}}+q$ by the same method adopted in the analysis of the correlation between $K_{\rm{sym}}$ and $L$. Taking $n=10^4$ independent runs of the simulation with each simulation $m=10^2$ points, we obtain $\overline{p}\approx 19.65\pm1.38$ and $\overline{q}\approx2550\pm342\,\rm{MeV}$. The large slope $\overline{p}$ means that the 
$J_{\rm{sym}}$ is very sensitive to the change in $K_{\rm{sym}}$. For example, $J_{\rm{sym}}$ is 979\,MeV (193\,MeV) if $K_{\rm{sym}}$ is taken as $-80\,\rm{MeV}$ ($-120\,\rm{MeV}$).
Thus, although the change in $K_{\rm{sym}}$ is only 40 MeV, the corresponding change in $J_{\rm{sym}}$ is about $-786\,\rm{MeV}$. This shows again obtaining the constraint on $J_{\rm{sym}}$ is very difficult unless the $K_{\rm{sym}}$ is very well constrained. It is interesting to note that the strong positive correlation between $J_{\rm{sym}}$ and $K_{\rm{sym}}$ was also reported in Ref.\,\cite{Mar19} (see Fig.\,7 there).

\section{Summary}\label{S8}
In summary, the unbound nature of PNM requires a sum rule linking intrinsically the ANM EOS parameters independent of any theory. 
By analyzing this sum rule at different orders in $\delta^2$, we found several novel correlations relating the characteristics of SNM EOS with those of nuclear symmetry energy. In particular, at the lowest-order of approximations, 
the bulk parts of the slope $L$, curvature $K_{\rm{sym}}$ and skewness $J_{\rm{sym}}$ of the symmetry energy are found to be $L\approx K_0/3, K_{\rm{sym}}\approx LJ_0/2K_0$ and $J_{\rm{sym}}\approx I_0L/3K_0$, respectively. 
High-order corrections to these simple relations can be written in terms of the small ratios of high-order EOS parameters. The resulting intrinsic correlations among the magnitude $S$, slope $L$, curvature $K_{\rm{sym}}$ and skewness $J_{\rm{sym}}$ of the nuclear symmetry energy reproduce very nicely their empirical correlations from various microscopic nuclear many-body theories and phenomenological models in the literature. 

The unbound nature of PNM is fundamental and the required intrinsic correlations among the characteristics of ANM are general. Since the EOS of PNM is the sum of two sectors: the SNM EOS and different orders of nuclear symmetry energy from expanding the ANM EOS $E(\rho,\delta)$ in even powers of $\delta$, the vanishing pressure of PNM at zero density naturally relates the characteristics of the two sectors. While much progress has been made by the nuclear physics community in probing separately characteristics of the two parts of the ANM EOS, very little is known about the correlations between the characteristics of SNM and those characterizing the symmetry energy. The intrinsic correlations among the characteristics of ANM EOS provide a novel and model-independent tool not only for consistency checks but also for investigating the poorly known high-density properties of neutron-rich matter by using those with smaller uncertainties.

\section*{Acknowledgement}
This work is supported in part by the U.S. Department of Energy, Office of Science, under Award Number DE-SC0013702, the CUSTIPEN (China-U.S. Theory Institute for Physics with Exotic Nuclei) under the US Department of Energy Grant No. DE-SC0009971.


\begin{thebibliography}{99}
%\begin{references}
\bibitem{Rin80} P. Ring and P. Schuck, \textit{The Nuclear Many Body
Problem}, Springer, 1980.

\bibitem{Dan02} P. Danielewicz, R. Lacey, and W.G. Lynch, Science,
\textbf{298}, 1592 (2002).

\bibitem{ditoro} V. Baran, M. Colonna, V. Greco, M. Di Toro, Phys. Rep. \textbf{410}, 335 (2005).

\bibitem{LCK08} B.A. Li, L.W. Chen, and C.M. Ko, Phys. Rep. \textbf{464}, 113 (2008).

\bibitem{Tesym} B.A. Li, \`{A}. Ramos, G. Verde, I. Vida\~{n}a (Eds.), \textit{Topical issue on nuclear symmetry energy}, Euro. Phys. J. A \textbf{50}, No.2 (2014).


\bibitem{Col14} G. Colo, U. Garg, and H. Sagawa, Eur. Phys. J. A \textbf{50}, 26 (2014).

\bibitem{Bal16} M. Baldo, G.F. Burgio, Prog. Part. Nucl. Phys. \textbf{91}, 203 (2016).

\bibitem{Oer17} M. Oertel, M. Hempel, T. Kl\"ahn, and S. Typel, Rev. Mod. Phys. \textbf{89}, 015007 (2017).

\bibitem{Garg18} U. Garg, G. Col\`{o}, Prog. Part. Nucl. Phys. \textbf{101}, 55 (2018).

\bibitem{Heb15} K. Hebler, J.D. Holt, J. Menendez, and A. Schwenk, Annu. Rev. Nucl. Part. Sci. \textbf{65}, 457 (2015).

\bibitem{Car15} J. Carlson, S. Gandolfi, F. Pederiva, Steven C. Pieper, R. Schiavilla, K. E. Schmidt, and R. B. Wiringa,
Rev. Mod. Phys. \textbf{87}, 1067 (2015).

\bibitem{FP} B. Friedman and V.J. Pandharipande, Nucl. Phys. {\bf A361}, 502 (1981).

\bibitem{MS} W. D. Myers and W. J. Swiatecki,  Acta Phys. Pol. B 26, 111 (1995)

\bibitem{Sch05} A. Schwenk and C.J. Pethick, Phys. Rev. Lett.
\textbf{95}, 160401 (2005).

\bibitem{Gez10} A. Gezerlis and J. Carlson, Phys. Rev. C \textbf{81}, 025803 (2010).

\bibitem{Hut20}S. Huth, C. Wellenhofer, and A. Schwenk, Phys. Rev. C {\bf 103}, 025803 (2021).

\bibitem{UG} M. J. H. Ku, A. T. Sommer, L. W. Cheuk, and M. W. Zwierlein, Science \textbf{335}, 563 (2012).


\bibitem{Tew17}I. Tews,  J.M. Lattimer, A. Ohnishi, and E.E. Kolomeitsev, 
ApJ. \textbf{848}, 105 (2017).


\bibitem{Gio08} S. Giorgini, L.P. Pitaevskii, and S. Stringari, Rev. Mod. Phys.
\textbf{85}, 1225 (2008).

\bibitem{Fa1} F. J. Fattoyev, C. J. Horowitz, J. Piekarewicz, and G. Shen, Phys. Rev. C \textbf{82}, 055803 (2010).

\bibitem{Fa2} F. J. Fattoyev, W. G. Newton, Jun Xu, Bao-An Li, Physical Review C \textbf{86}, 025804 (2012).

\bibitem{PPNP} B.A. Li, B.J. Cai, L.W. Chen, and J. Xu, Prog. Part. Nucl. Phys. \textbf{99}, 29 (2018).


\bibitem{Newton12} W.G. Newton, M. Gearheart, and B.A. Li, Astrophys. J. Supplement Series \textbf{204}, 9 (2013).

\bibitem{Newton20} W.G. Newton and G. Crocombe, arXiv:2008.00042 (2020).

\bibitem{Mar19}J. Margueron and F.  Gulminelli, Phys. Rev. C \textbf{99}, 025806 (2019).

\bibitem{Maz13} X. Roca-Maza et al., Phys. Rev. C \textbf{88}, 024316 (2013).

\bibitem{Lida14} K. Lida, K. Oyamatsu, Euro. Phys. J. A \textbf{50}, 42 (2014).

\bibitem{Pro14} C. Provid\^{e}ncia \textit{et al.}, Euro. Phys. J. A \textbf{50}, 44 (2014).

\bibitem{Mon17} C. Mondal, B. K. Agrawal, J. N. De, S. K. Samaddar, M. Centelles, and X. Viñas, Phys. Rev. C \textbf{96}, 021302(R) (2017).


\bibitem{Hol18} J.W. Holt and Y. Lim, Phys. Lett. \textbf{B784}, 77 (2018).


\bibitem{LiBA2020}B.A. Li and M. Magno, Phys. Rev. C \textbf{102}, 045807 (2020).

\bibitem{Han20} H. Gil, Y.M. Kim, P. Papakonstantinou, and C.H. Hyun, arXiv:2010.13354 (2020).


\bibitem{Mal20} T. Malik, B.K. Agrawal, C. Provid\^{e}ncia and J.N. De, Phys. Rev. C \textbf{102}, 052801(R) (2020).

\bibitem{Mal18}T. Malik, N. Alam, M. Fortin, C. Provid\^{e}ncia, B. K. Agrawal, T. K. Jha, Bharat Kumar, and S. K. Patra, Phys. Rev. C \textbf{98}, 035804 (2018).

\bibitem{Li19} B.A. Li, P.G. Krastev, D.H. Wen and N.B. Zhang, Eur. Phys. J. A \textbf{55}, 39 (2019).


\bibitem{Cai17x}B.J. Cai and L.W. Chen, Nucl. Sci. Tech. \textbf{28}, 185 (2017).


\bibitem{Che09} L.W. Chen, B.J. Cai, C.M. Ko, B.A. Li, C. Shen, and
J. Xu, Phys. Rev. C \textbf{80}, 014332 (2009).

\bibitem{LiBA13} B.A. Li and X. Han, Phys. Lett. \textbf{B727}, 276 (2013).


\bibitem{Ohio20}C. Drischler, R.J. Furnstahl, J.A. Melendez and D.R. Phillips, Phys. Rev. Lett. {\bf 125}, 202702 (2020).


\bibitem{You99}D.H. Youngblood, H.L. Clark, and Y.-W. Lui, Phys. Rev. Lett. \textbf{82}, 691 (1999).
\bibitem{Shl06}S. Shlomo, V.M. Kolomietz, and G. Colo, Eur. Phys. J. A \textbf{30}, 23 (2006).
\bibitem{Che12} L.W. Chen and J.Z. Gu, J. Phys. G \textbf{39}, 035104 (2012).

\bibitem{XieLi-JPG}W.J. Xie and B.A. Li,  J. Phys. G: Nucl. Part. Phys. {\bf 48}, 025110 (2021).

\bibitem{Xie20}W.J. Xie and B.A. Li, ApJ. \textbf{899}, 4 (2020).


\bibitem{LiBA98} B.A. Li, C.M. Ko, and  W. Bauer, Int. J. Mod. Phys. E
\textbf{7}, 147 (1998).
\bibitem{Cai12} B.J. Cai and L.W. Chen, Phys. Rev. C \textbf{85}, 024302 (2012).
\bibitem{Gon17} C. Gonzalez-Boquera, M. Centelles, X. Vinas, and A. Rios, Phys. Rev. C \textbf{96}, 065806 (2017).
\bibitem{PuJ17} J. Pu, Z. Zhang, and L.W. Chen, Phys. Rev. C \textbf{96}, 054311 (2017).

\bibitem{Lee98} C.H. Lee, T.T. S. Kuo, G.Q. Li, and G.E. Brown, Phys. Rev. C \textbf{57}, 3488 (1998).


\bibitem{Rus16}P. Russotto \textit{et al.}, Phys. Rev. C \textbf{94}, 034608 (2016).


\bibitem{Xie19}W.J. Xie and B.A. Li, ApJ. \textbf{833}, 174 (2019).


\bibitem{Blu20}A. Blum, J. Hopcroft, and R. Kannan, \textit{Foundations of Data Science}, Cambridge University Press, 2020, Chap.1.

\bibitem{Som20}R. Somasundaram, C. Drischler,
I. Tews, and J. Margueron, arXiv:2009.04737 (2020).

%\end{references}
\end{thebibliography}
\end{document}